\documentclass[aps,pre,twocolumn,showpacs,superscriptaddress,nofootinbib]{revtex4}

\usepackage{graphicx}
\usepackage{dcolumn}
\usepackage{bm}

\def\be{\begin{equation}}
\def\ee{\end{equation}}
\def\bea{\begin{eqnarray}}
\def\eea{\end{eqnarray}}
\def\bi{\begin{itemize}}
\def\ei{\end{itemize}}

\begin{document}
\preprint{\today}


\title{ Linear Temperature-Mass relation and Local virial relation:
  Two hypotheses for self-gravitating systems}

\author{Yasuhide Sota \footnote{sota@cosmos.phys.ocha.ac.jp}}
  \affiliation{Department of Physics, Ochanomizu University,
	       2-1-1 Ohtuka, Bunkyo, Tokyo 112-8610, Japan}
  \affiliation{Advanced Research Institute for Science and Engineering, 
               Waseda University, Ohkubo, Shinjuku--ku, Tokyo 169-8555, Japan}

\author{Osamu Iguchi \footnote{osamu@phys.ocha.ac.jp}}
  \affiliation{Department of Physics, Ochanomizu University,
	       2-1-1 Ohtuka, Bunkyo, Tokyo 112-8610, Japan}

\author{Akika Nakamichi \footnote{akika@astron.pref.gunma.jp}}
  \affiliation{Gunma Astronomical Observatory,
               6860-86, Nakayama, Takayama, Agatsuma, Gunma 377-0702,
  Japan}

\author{Masahiro Morikawa \footnote{hiro@phys.ocha.ac.jp}}
  \affiliation{Department of Physics, Ochanomizu University,
               2-1-1 Ohtuka, Bunkyo, Tokyo 112-8610, Japan}

\begin{abstract}
We propose two hypotheses which characterize the 
quasi-equilibrium state
that realizes 
after a cold collapse of self-gravitating system. The first
hypothesis is the linear temperature-mass (TM) relation, which yields a
characteristic  non-Gaussian velocity distribution. The second  is
the local virial (LV) relation, which, combining the linear TM relation,
determines a unique mass density profile as $\rho(r)={\rho }%
_{0}r^{-4}e^{-r_0/r}$. Although this density profile is unphysical
in the central region, the region is just inner a few  percent 
around the center of a bound
region in cumulative mass, which is beyond
the resolution of our numerical simulations. Hence posing two
hypotheses is compatible to the numerical simulations
for almost the whole region of the virialized bound state. 
Actually, except for this inner part, 
 this density profile fits well to  the data of
cold collapse simulations.  Two families
of spherical and isotropic models, polytropes  and King models, are examined
from a view point of these two hypotheses. We found that the LV relation
imposes a strong constraint on these models: only polytropes with index $n \sim
5$ such as Plummer's model
are compatible with the numerical results characterized by the two
hypotheses among these models. King models with the concentration parameter $%
c \sim 2$ violate the LV relation while they are consistent with the $R^{1/4}
$ law for the surface brightness. Hence the above characteristics can serve
as a guideline to build up the models for the bound state after a cold
collapse, besides the conventional criteria concerning the asymptotic
behavior.
\end{abstract}




\maketitle


\section{Introduction}

Collisionless self-gravitating systems (SGS) eventually settle down to a
quasi-equilibrium state through the phase mixing and the violent relaxation
processes under the potential oscillation\cite{Lynden67}. This quasi-equilibrium state is a prototype of the astronomical
objects such as galaxies and clusters of galaxies. These objects often show
universal profiles in various aspects.

It would be true that the energy distribution function and the mass density
distribution function are quite appropriate to characterize SGS, and actually
many studies have been done from this viewpoint. For example, the density
profile is found to be $\rho \propto r^{-4}$ in spherical and cold
collapses, \textit{i.e.,} collapses with small initial virial ratio 
\cite{Henon64,Albada82} and $\rho \propto r^{-3}-r^{-4}$ in the cluster-pair
merging processes\cite{Merrall03}.

On the other hand, the velocity distribution function has not fully been
examined so far despite its importance \footnote{%
So far, full velocity distributions have been mainly analyzed only for the
line-of-sight velocity profiles because they can be directly compared with
the observational data \cite{Dejonghe87,Gerhard93}}; anisotropy of
velocity dispersion or lower moments of velocity distributions with respect
to the deviations from Gaussian distributions have been mainly examined
(e.g. \cite{Binney82,Gerhard93}). However, the full velocity distribution
function plays an important role in identifying the quasi-equilibrium state,
which should be properly characterized by the whole phase-space probability
functions. Merrall \textit{et al.} numerically showed that the velocity
distributions in the central regions of the bound particles become Gaussian,
when the bound particles are well relaxed \cite{Merrall03}. However, since
SGS are manifestly non-additive systems \cite{Nakamichi04}, the ordinary
Gaussian distribution would not be expected for the whole bound particles.
Moreover, such non-Gaussian velocity distributions would be firmly connected
with the density profile characteristic to SGS %
\cite{Gerhard93,Kazantzidis04}.

Recently, we have simulated self-gravitating N-body systems by utilizing
the leap-frog symplectic integrator on GRAPE-5, a special-purpose computer
designed to accelerate N-body simulations\cite{GRAPE}. 
The simulations were performed mainly 
in the case with cold collapse initial conditions. 
It is certain that the processes may be too simplified 
to deduce the ultimate nature of actual SGSs, 
in which merging or tidal effects take place.
However, one of our main goals is to establish 
the universal properties of SGS through the violent relaxation process, 
i.e. the relaxation through the oscillation of gravitational potential. 
This process is rather common in actual SGSs 
such as galaxies and dark matters. 
In this context, the cold collapse is considered to be a typical case 
which causes the violent relaxation. 
Hence it is meaningful to examine the cold collapse as a first step.

When the system
experiences violent gravitational processes such as a cold collapse and
a cluster-pair collision, we obtained a universal velocity distribution
profile expressed as the democratic (=equally weighted) superposition of
Gaussian distributions of various temperatures(DT distribution
hereafter) 
in which the local temperature $T(r)$ is defined 
using the local velocity variance $\langle v^{2}\rangle $ as 
$T(r):=m\langle v^{2}\rangle (r)/3k_{B}$, where $m$ is the particle mass
\cite{Osamu04}. Moreover, we have found that the locally defined
temperature linearly falls down in the intermediate cluster region outside
the central part, provided it is described against the cumulative mass $M_{r}
$, \textit{i.e.,} $dT/dM_{r}=const.$ This fact is consistent with the
appearance of DT distributions.

Furthermore to the linear TM relation, we have also obtained another
peculiar fact that the LV relation between the locally defined
potential energy and kinetic energy holds except for the weak fluctuations.
More precisely, the local temperature $T(r)$ is proportional to the local
potential $\Phi (r)$, with constant proportionality $6k_{B}T(r)=-m\Phi (r)$
for a wide class of cold collapse simulations. 
It was originally pointed out by Eddington
that Plummer's model
satisfies this condition exactly  \cite{Eddington16}.
 Moreover it was also pointed
out that Plummer's model can be extended to 
the anisotropic model of a stellar
system satisfing the LV relation \cite{Evans05}.
We found that the bound region, \textit{i.e.,%
} the region composed of the bound particles 
keeps this relation quite well at least during the time interval much longer
than the initial crossing time. 
This is a prominent correlation in the phase space distribution function,
and also constructs another backbone character of SGS after a cold collapse.

In sec.\ref{sec:level2} and \ref{sec:level3}, 
we will show that the above two hypotheses  for SGS 
are supported by the results of a variety of cold collapse simulations. 
Then in sec.\ref{sec:level4}, combining the two hypotheses, 
\textit{i.e.,} the linear TM relation and the LV relation, 
we obtain a unique density profile 
$\rho (r)=\rho_{0}(r/r_{0})^{-4}e^{-r_{0}/r}$ 
which asymptotically behaves as $\rho\propto r^{-4}$. 
Justification of this density profile will also be discussed. 
In sec.\ref{sec:relaxation}, 
we will ascertain that the above two hypotheses can arise even 
in the collisionless quasi-equilibrium state, 
where the region inside a half-mass radius is not yet 
under the effect of two-body relaxation. 
In sec.\ref{sec:models} 
the above two hypotheses will be explored 
in two typical collisionless static models with 
spherical configuration and isotropic velocity dispersions; 
polytropes and King models. 
Sec.\ref{sec:conclusions} is devoted to the summaries and
conclusions of this paper. 

\section{ Linear TM relation}

\label{sec:level2}

Let us now further consider the linear TM relation. In our recent work %
\cite{Osamu04}, we have examined the velocity distribution function of
collisionless SGS after a cold collapse in numerical N-body calculations.
According to this work, immediately after a cold collapse, well-relaxed and
almost spherical-symmetric bound state is formed. In order to examine the
local property, we use the cumulative mass $M_{r}$ within the radius $r$,
 \textit{i.e.,} $M_{r}:=4\pi \int_{0}^{r}dr^{\prime }r^{\prime 2}\rho
(r^{\prime })$ as a measure of the radial distance from the cluster center,
which is defined as the position of the potential minimum. In our numerical
simulation, we divide the whole system into several concentric shells with
equal width in the coordinate $M_{r}$ and consider the averaged quantities
within each shell as local variables. 

In our simulation, particles in the outermost region, which includes $20\%$
of the total mass, escape almost freely just after the cold collapse and are
irrelevant to the quasi-equilibrium state. Therefore we concentrate on the
local temperature within the inner region initially including $80\%$ of the
total mass. More precisely, we define the bound particles as the particles
inside the $M_r$, at which the local temperature $T(M_r)$ takes the minimum
value. Hereafter, we pay our attention to the bound region composed of these
bound particles. In the central part of this region, the two-body relaxation
proceeds and the particles evaporate toward outside. Moreover, in this
central region, the effect of cutoff $\epsilon$ proposed in numerical
simulations cannot be ignored completely. Discarding this part from the
bound region, we found, that the local velocity variance, \textit{i.e.,} the
local temperature linearly decreases to zero in the coordinate $M_{r}$, 
\begin{equation}
\langle v^{2}\rangle (r)=\frac{3k_{B}T}{m}=c(M_{tot}(t)-M_{r}),  
\label{TMrel}
\end{equation}
where $c$ is a positive constant and $M_{tot}(t)$ is the total mass of the
bound particles at $t$. $M_{tot}(t)$ is initially equal to $80\%$ of the
total mass but is gradually diminished, as several particles evaporate from
the bound state.

\begin{figure} 
\begin{center}
\includegraphics[width=8cm]{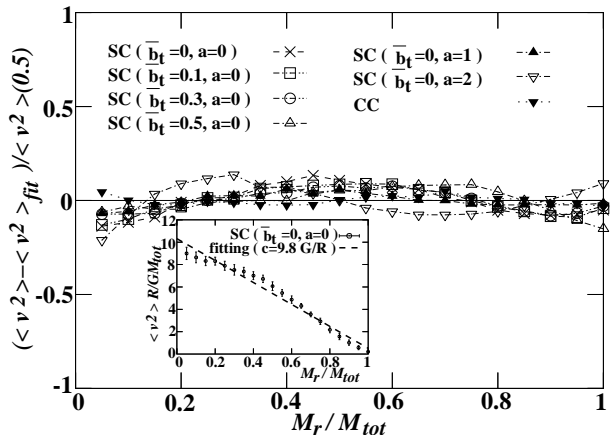}
\end{center}
\caption{ A mass dependence of velocity dispersion ($\langle
v^2(M_r/M_{tot}) \rangle$) obtained by a typical cold collapse simulation;
spherical collapse (SC) and cluster-pair collisions (CC). In the case of SC, 
$5000$ particles are distributed with a power law density profile ($\protect%
\rho\propto r^{-a}$) within a sphere of radius $R$ and the initial virial
ratio ($\bar{b}_t$) is set to be small. In the case of CC, each cluster has
the equal number of particles ($2500$) and all particles are homogeneously
distributed within a sphere of radius $R$ and is set to be virialized
initially. The initial separation of the pair is $6R$ along the $x$ axis. In
all of the simulations, softening length $\protect\epsilon=2^{-8}R$ is
introduced to reduce the numerical error caused by close encounters. The
inset figure shows that locally averaged velocity dispersion is plotted as a
function of $M_{r}(t)/M_{tot}(t)$ for SC($\bar{b}_t=0,a=0$) as a typical
example. Apparent linearity is remarkable. The broken line is the best fit
line with $c=9.8G/R$ (Eq.(\ref{TMrel})). The open circles represent the
time averaged numerical data from $t=5t_{ff}$ and $t=100t_{ff}$, where $%
t_{ff}$ is the initial free fall time defined as 
 $t_{ff}:=\protect\sqrt{%
R^{3}/GM_{tot}(0)}$. The deviation from the best fit line ($<v^2>_{fit}$)
normalized by the velocity dispersion at the half mass is shown for some
numerical simulations with different initial conditions. The value of each
deviation is under $10\%$. }
\label{TM}
\end{figure}

\begin{figure} 
\begin{center}
\includegraphics[width=8cm]{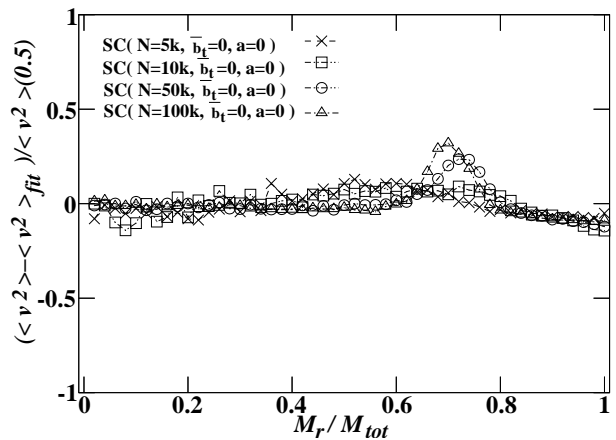}
\end{center}
\caption{ A mass dependence of velocity dispersion ($\langle
v^2(M_r/M_{tot}) \rangle$) obtained by a typical cold collapse simulation
SC ($\bar{b}_t=0,a=0$) with different particle number $N$.
The case with $N=5\times 10^3, 10^4, 5 \times 10^4, 10^5$ are depicted.
The peak appears in the case with $N= 5 \times 10^4, 10^5$ around
$M_r \sim 0.7 M_{tot}$, which represents
a shock region.
}
\label{TM-diffN}
\end{figure}

Here we examined the deviation of TM relation from the linear relation (\ref
{TMrel}) for a variety of simulations of cold collapse. For the estimate of
the deviation, we first take the time-average for the velocity dispersion at
each shell to derive the time-averaged $\langle v^{2}\rangle$-mass relation.
Then we linearly fit it as (\ref{TMrel}) to derive $\langle
v^{2}\rangle_{fit}$ and subtract it from $\langle v^{2}\rangle$. In Fig.\ref
{TM}, the deviation normalized by $\langle v^{2}\rangle$ at the half mass
radius are depicted for several simulations with  total particle's number
$N=5000$. The deviations are very small
and are less than 10 percent at each $M_r$ for all of the simulations we
examined. Hence, this result is quite robust against initial conditions
with $N=5000$
provided the collapse is cold. Even for the case of cluster-pair collision,
this linear relation is widely obtained.
We also examined the $N$ dependence of the linear TM relation for the
case with initially homogeneous spherical sphere
(Fig.\ref{TM-diffN}). As $N$ increases, the particles collapse more and 
more coherently, which causes a shock region moving
slowly outward after a collapse.
Although such a shock region enhances the deviation from the 
linear TM relation, 
the region within a shock region satisfies the linearity quite well.

\section{ Local virial relation}

\label{sec:level3} 
Besides the linear TM relation, there is another characteristic for
SGS as we now argue. It is well known that the gravitationally bound system
approaches a virialized state satisfying the condition 
\begin{equation}
\overline{W}+2\overline{K}=0,  \label{Virial}
\end{equation}
where $\overline{W}$ and $\overline{K}$ are, respectively, the averaged
potential energy and kinetic energy of the whole bound system. This is a
global relation that holds for the entire system, after the initial coherent
motion fades out.

Here we define the locally averaged potential energy and kinetic energy
inside the radius $r$, respectively as

\begin{eqnarray}
\overline{W}_{r} &:=& \frac{1}{2}\int_{0}^{r}\Phi (r^{\prime })
\rho(r^{\prime })4\pi r^{\prime 2}dr^{\prime },  \label{poteall} \\
\overline{K}_{r} &:=& \int_{0}^{r}\frac{\langle v^{2}\rangle (r^{\prime})}{2}
\rho (r^{\prime })4\pi r^{\prime 2}dr^{\prime },  \label{kineall}
\end{eqnarray}
where $\star (r^{\prime })$ means the local object $\star $ evaluated at $%
r^{\prime }$.

Then we extend the virial relation (\ref{Virial}) locally as 
\begin{equation}
\overline{W}_{r}+2\overline{K}_{r}=0,  \label{localvir}
\end{equation}
and examine how precisely this relation is locally attained inside a bound
state.

\begin{figure} 
\begin{center}
\includegraphics[width=8cm]{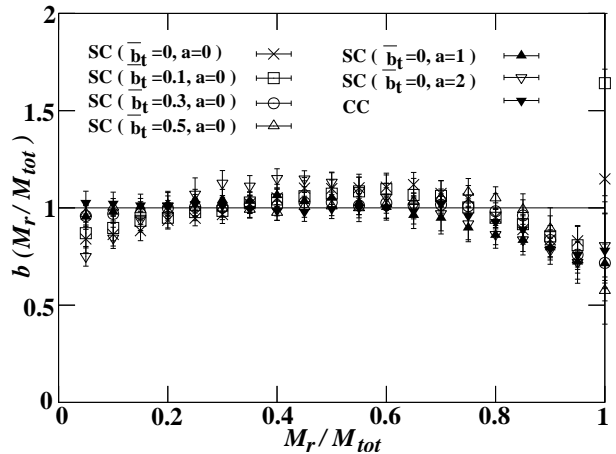}
\end{center}
\caption{ The LV relation for some numerical simulations with N=$5000$. 
The local virial ratio $b$ is plotted as a function of $%
M_{r}/M_{tot}$. The initial condition of each simulation is the same as Fig.%
\ref{TM}. The virial ratios of each shell are time averaged from $t=5t_{ff}$
and $t=100t_{ff}$. }
\label{LV}
\end{figure}

\begin{figure} 
\begin{center}
\includegraphics[width=8cm]{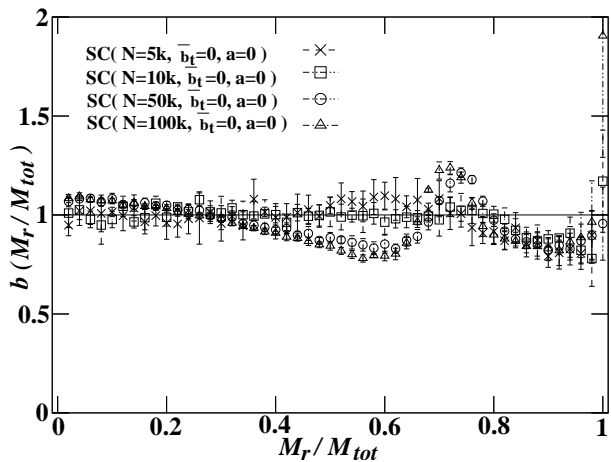}
\end{center}
\caption{ The LV relation for some numerical simulations with different
particle number $N$. The local virial ratio $b$ is plotted as a function of $%
M_{r}/M_{tot}$.The case with $N=5\times 10^3, 10^4, 5 \times 10^4, 10^5$ are depicted. The initial condition of each simulation is the same as Fig.%
\ref{TM-diffN}. The virial ratios of each shell are time averaged from $t=5t_{ff}$
and $t=100t_{ff}$. The peak appears in the case with $N= 5 \times 10^4, 10^5$ around
$M_r \sim 0.7 M_{tot}$, which represents
a shock region.}
\label{LV-diffN}
\end{figure}

More precisely, the above relation Eq.(\ref{localvir}) is equivalent to the
purely local relation 
\begin{equation}
2\langle v^{2}\rangle (r)=-\Phi (r),  \label{localvir2}
\end{equation}
at each position $r$. Hence, we define the LV ratio 
\begin{equation}
b(r) := -2 \langle v^{2}\rangle (r)/\Phi(r) ,  \label{localvirr}
\end{equation}
and examine the value of $b(r)$ for each shell. For a wide class of
collapses including cluster-pair collision examined in \cite{Osamu04}, the
value $b(r)$ takes almost unity: it deviates from unity less than 10 percent
upward in the intermediate region and a little more downward in the inner
and outer region for all of the simulations with $N=5000$(Fig.\ref{LV}). Hence a wide
class of cold collapse simulations with $N=5000$ yield the relation (\ref{localvir2})
quite well. We also examined the N-dependence of the LV relation for the
cases in Fig.\ref{LV-diffN}. We got the result that
the LV  relation is affected by the shock as well as the linear
TM relation in the case with $N \geq 5 \times 10^4$,where
the LV ratio $b$ takes the peak at the shock region.
the value $b$ becomes smaller than unity around the shock region in order
to compensate for the enhance at the shock region.
However,  $b$ oscillates  around  unity even in such cases, which
supports the LV relation in the zeroth order.  
Hence,this LV relation should be an another  characteristic of
SGS, and this is our second hypothesis.


\section{Density profile}

\label{sec:level4} 
We have explored the two hypotheses of SGS so far, \textit{i.e.,} the linear
TM relation Eq.(\ref{TMrel}) and the LV relation Eq.(\ref{localvir2}) for
the bound cluster. Combining these two relations, we now derive a mass
density profile. We simply substitute Eq.(\ref{localvir2}) into Eq.(\ref
{TMrel}) to obtain $2c(M_{tot}-M_{r})=-\Phi $. Differentiating this with $r$
and using Poisson's equation, we obtain a differential equation for the mass
density $\rho (r)$: 
\begin{equation}
\frac{d}{dr}(r^{4}\rho )=\frac{G}{2c}r^{2}\rho ,
\end{equation}
which admits the unique solution 
\begin{equation}
\rho (r)={\rho }_{0}\left(\frac{ r_0 }{r}\right)^4e^{-r_0/r },
\label{sotakai}
\end{equation}
where $r_{0}:=G/2c$ and $\rho _{0}$ is a constant. 
The cumulative mass $M_r$ for this density profile is easily derived as
\begin{equation}
M_r=M_{tot}e^{-r_0/r },
\label{sotakaicum}
\end{equation}
where $M_{tot}$ is the total mass and is described as 
\be
M_{tot}  = 4\pi r_0^3 \rho _0.
\label{sotakaimtot}
\ee

 The central part of this density is unphysical, since it
increases with increasing radius $r$ inside $r_{\ast} := r_0/4$. This fact
warns that the two hypotheses we proposed are not always justified in
the entire region. However, $M_r$ inside e $r_{\ast}$ is just
\be
M_r(r_{\ast})=e^{-4}M_{tot}\sim0.0183M_{tot},
\ee
from (\ref{sotakaicum}). Hence this unphysical region is at most 2 percent
of the full bound region in cumulative mass. 

\begin{figure} 
\begin{center}
\includegraphics[width=8cm]{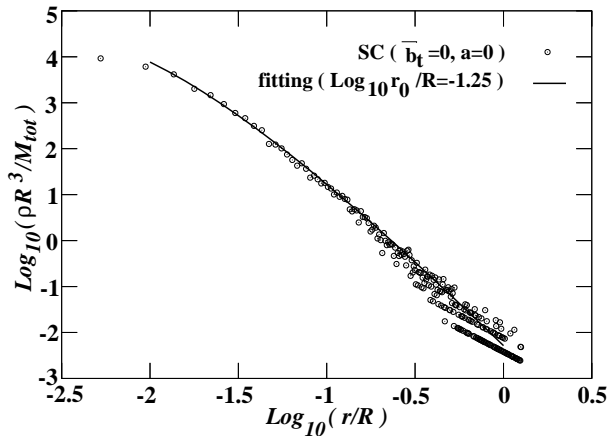}
\end{center}
\caption{ A log-log plot of a typical density profile after a cold collapse.
The initial condition of this simulation is the same as SC $(\bar{b}_t=0,
a=0)$ in Fig.\ref{TM}. The open circles represent the numerical data at $%
t=100t_{ff}$. The solid line is the density profile (Eq.(\ref{sotakai}))
with the best fit parameter ($\mbox{Log}_{10}(r_0/R)=-1.25$), which
corresponds to $M_r|_{r=r_0}/M_{tot}\sim 0.43$. }
\label{sbfit-sim}
\end{figure}

Even if we put a constant density at this inner unphysical 
region and continuously connect it to the outer analytical 
solution (\ref{sotakai}), this inner  region is at most three percent
of the full bound region in mass (See Appendix\ref{appjoin}),
which is inner than the smallest value of $M_r$ for numerical data points
in Fig.\ref{TM} and Fig.\ref{LV}.
Hence we can safely neglect this unphysical region. 
Actually the function (\ref{sotakai}) fits well to 
our numerical results (Fig.\ref{sbfit-sim}). The asymptotic behavior $\rho
(r) \propto r^{-4}$ is consistent with previous works \cite{Jaffe87,makino90}. In sec.\ref{sec:models}, we will see how this central part is
modified with several collisionless static solutions.

\section{Collisionless or collisional? 
the properties of the system compatible with the two hypotheses}
\label{sec:relaxation}
In the previous section, 
we numerically showed that the bound system
after a cold collapse retains the two hypotheses 
during a long time interval on average.
Here a natural question arises, \textit{i.e.,} 
do these two characteristics occur just after a cold collapse
or are they gradually developed through the two-body relaxation?
Now we check this issue by estimating 
the two-body relaxation time scale $t_{rel}$, which is described
as
 \be
t_{rel} = 0.065\frac{{\overline{
\langle v^{2} \rangle}^{3/2}}}
 {{m G^2 \overline{\langle \rho \rangle}\log \left( {1/\epsilon } \right)}},
\label{trel}
 \ee
where
$\overline{\langle v^{2} \rangle}$ and
$\overline{\langle \rho \rangle}$
are the velocity variance and the density averaged 
inside the half mass radius, respectively \cite{Spitzer87}.
Although  $t_{rel}$ is a function of time $t$,  the value
settles down to   a  constant value  $t^{\ast}_{rel}$ when
the system stays in a  quasi-equilibrium bound state.  
 From the definition of $t_{rel}(t)$, the system
inside a half mass radius
naively attains two-body relaxation at $t>t^{\ast}_{rel}$.
In Table.\ref{tab:tr}, we show the value of  $t^{\ast}_{rel}$ for
several simulations.

For example, 
because of the high concentration of the bound particles
just after a collapse,
 $t^{\ast}_{rel}$ becomes very short  in the case with 
a cold collapse from a homogeneous sphere
(run SC($N=5000$ and $50000,\overline{b_t}=0,a=0 $)).
In this case, we cannot regard the bound state after a collapse
as a collisionless system, since $t^{\ast}_{rel}$ is so short
that the two-body relaxation is 
rapidly  achieved just after a cold collapse.
On the other hand,  $t^{\ast}_{rel}$ becomes longer
by increasing the initial virial ratio or by changing the initial
particle configuration to be inhomogeneous. 
Especially, 
in the case of a mild collapse with initial virial ratio $0.5$ 
(run SC($N=5000,\overline{b_t}=0.5,a=0 $)),
 $t^{\ast}_{rel}$ becomes
as long as $40t_{ff}$.

\begin{table}
\caption{\label{tab:tr}
The relaxation time  for a bound state $t^{\ast}_{rel}$ is depicted.
The initial condition of each simulation is 
a spherical collapse (SC).
Each parameter $N$, $\overline{b_t}$, and $a$ denote 
the particle number, the initial  virial ratio, 
and the initial density profile $\rho\sim r^{-a}$ respectively.
}
\begin{center}
\begin{tabular}{lr}\hline
run & $t^{\ast}_{rel}$ \\\hline\hline
SC($N=5000, \overline{b_t}=0, a=0$)   & $3t_{ff}$  \\
SC($N=50000, \overline{b_t}=0, a=0$)  & $13t_{ff}$ \\
SC($N=5000, \overline{b_t}=0.5, a=0$) & $41t_{ff}$ \\
SC($N=5000, \overline{b_t}=0, a=2$)   & $24t_{ff}$ \\\hline
\end{tabular}
\end{center}
\end{table}%

%

\begin{figure} 
\begin{center}
\includegraphics[width=8cm]{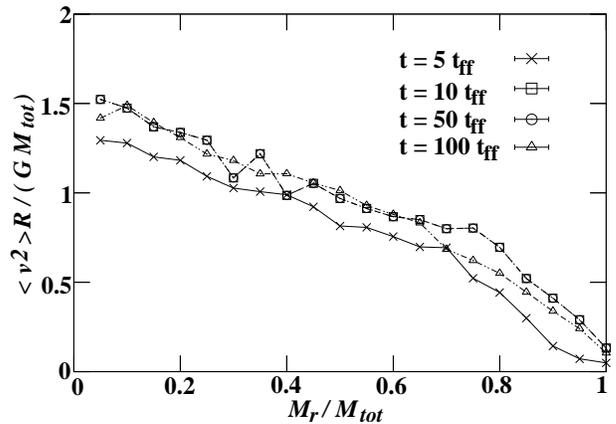}
\end{center}
\caption{A mass dependence of velocity dispersion ($<v^2(M_r)>$) 
for the case of a spherical collapse (run SC($N=5000,\overline{b_t}=0.5,a=0$)),
for which  $t^{\ast}_{rel}$ is almost $41t_{ff}$.
The velocity dispersion of each shell at the different time 
($t=5t_{ff}$, $10t_{ff}$, $50t_{ff}$, and $100t_{ff}$) are superposed.
}
\label{N5kv0_TM}
\end{figure}

Here with this mild collapsing case, 
we examined if our two hypotheses are achieved 
even before the two-body relaxation is developed in the bound state. 
In the case with the linear TM relation, 
the slope keeps almost constant 
even at the time in which the two-body relaxation has not been attained
(Fig.\ref{N5kv0_TM}) except for the slight shift of the slope 
in the outer part at $M_r \sim 0.7 M_{tot}$.
This constant slope is conserved 
even after the half of the bound state are 
under the effect of two-body relaxation ($t\geq 41 t_{ff}$). 
Hence the character of the linear TM relation appears during
the period that the system is almost collisionless and
is conserved  even after the two-body relaxation
is attained in more than half of the full bound region.  

\begin{figure} 
\begin{center}
\includegraphics[width=8cm]{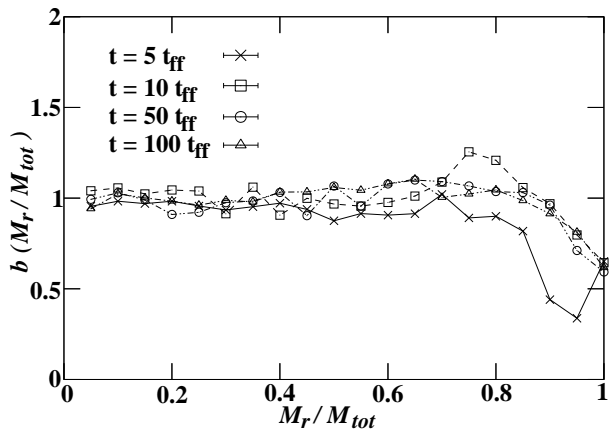}
\end{center}
\caption{The local virial relation for run SC($N=5000,\overline{b_t}=0.5,a=0 $)
for which $t^{\ast}_{rel}$ is almost $41t_{ff}$.
The ratio $b(M_r/M_{tot})$ is plotted as a function of $M_{r}$. 
The virial ratios of each shell at the different time 
($t=5t_{ff}$, $10t_{ff}$, $50t_{ff}$, and $100t_{ff}$) are superposed.
}
\label{N5kv0_LV}
\end{figure}

The LV  relation $b(M_r) = 1$ is also attained 
even in the early stage,
except for larger fluctuations in the outer
part $M_{r}> 0.7M_{tot}$ (Fig.\ref{N5kv0_LV}), although this deviation
has nothing to do with the two-body relaxation.
Since the half of the bound state is not relaxed 
at $t=10t_{ff}$ in Table.\ref{tab:tr},
we can claim that
the LV relation is attained quite well even in the collisionless region.

In summary, both the linear TM and the LV relation  appear at
the stage in which the system can be regarded as collisionless, although
in the outer part the slope of TM is
slightly shifted and $b(M_r)$ fluctuates a little larger. 
The fluctuation occurs at almost the same region as the one where the slope
slightly shifts in TM relation. We ascertained that
the velocity dispersion is anisotropic
at almost  the same region in these simulations.
Hence, we suspect that both  the shift of the slope and the large fluctuation
in the outer part are closely related to the 
anisotropy of the velocity dispersion
in this region. 
 In the next section, on the basis of  these results
we will examine two typical collisionless static models of
SGS.


\section{Comparison with Spherical and isotropic static models}

\label{sec:models} 
We have extracted the essence of cold collapses of SGS in the form of two
hypotheses, \textit{i.e.,} the linear TM relation Eq.(\ref{TMrel}) and the
LV relation Eq.(\ref{localvir2}). How are they justified?

A collisionless SGS can be described by the Vlasov
equation for the phase space distribution function $f(\vec{r},\vec{v})$ in
the mean field limit. Especially any function of the integral of motion can
be a stationary solution of the Vlasov equation. Here we treat several
collisionless static and spherical solutions of SGS and examine which class of
solutions meet the two hypotheses we proposed.

In a cold collapse, the velocity dispersion of a bound
system becomes radially anisotropic especially in the outer regions %
\cite{Albada82}. In the preceding  section, however, we showed that the
two hypotheses are achieved quite well even in the inner isotropic
and collisionless region for a mild collapsing case. 
Hence, as zero-th order description, we regard the bound state as an
isotropic system  and compare it with the spherical and static model with an
isotropic velocity dispersion described as $f(E)$, in terms of the particle
energy per unit mass $E(r,v):=\Phi (r)+v^{2}/2$. Here we pay our attention
to two sorts of solutions, polytropes and King models as the examples of
those models with a flat boundary condition at the center.
We  compare them with the typical numerical data which
meet both of our two hypotheses, that is, SC ($\bar{b}_t=0,a=0$) with  $N=10^4$.

First, we examined polytropes. Polytropes are the distribution functions
with the form $f(r,v)=\tilde{E} (r,v)^{n-3/2}$, where $\epsilon $ is the
relative energy defined as $\tilde{E} :=-E+\Phi _{\ast }$, $\Phi _{\ast }$ is
chosen so as to guarantee $f>0$ for $\tilde{E} >0$ and $n$ is the polytropic
index \cite{Binney87}. 
Polytropes with $n>5$
necessitates a boundary to make the total mass finite since it diverges at
infinity. Here, since we are interested in the case with no boundary, we
focus our attention to the case with $n\leq 5$.

For polytropes with $n>1/2$, it is easy to find that the local temperature $%
T(r)$ is connected to the local potential $\Phi (r)$ as, 
\begin{equation}
\langle v^{2}\rangle (r)=\frac{3k_{B}T}{m}=\frac{3}{1+n}(\Phi _{\ast }-\Phi
(r)),
\end{equation}
which is equivalent to 
\begin{equation}
\frac{2\overline{K}_{r}}{\overline{W}_{r}}=\frac{6}{{n+1}}\left( {\frac{{%
M_{r}}}{2\overline{W}_{r}}\Phi _{\ast }-1}\right) .
\end{equation}
(See Appendix \ref{ap-LV} for the derivation.) Since $\Phi _{\ast }=0$ for $%
n=5$, the LV relation Eq.(\ref{localvir}) or (\ref{localvir2}) is attained
for $n=5$, which means that the LV ratio (\ref{localvirr}) becomes unity for 
$n=5$. This special solution with $n=5$ can be analytically expressed as (%
\ref{phi_plimmer}) and is known as Plummer's model \cite{Binney87}.

\begin{figure} 
\begin{center}
\includegraphics[width=8cm]{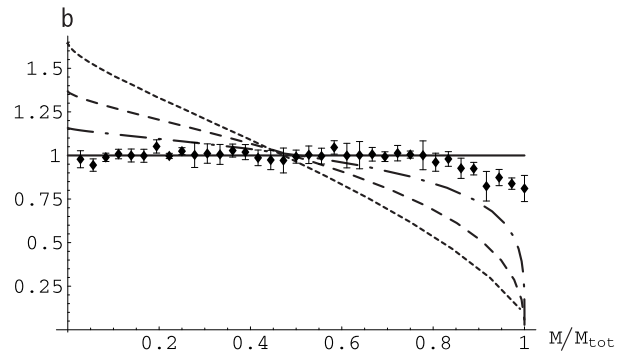}
\end{center}
\caption{ LV relation for polytropes with the unit of $G=r_h=M_{tot}=1$. The solid curve refers to the LV ratio 
$b(M_r/M_{tot})$ as a function of cumulative mass $M_r/M_{tot}$ for Plummer's model.
The other curves refer to $b(M_r/M_{tot})$ for polytropes with $n=3.0$ (dot-dashed), $n=1.5
$ (dashed), $n=0.51$ (dotted). The plots represent the numerical data of SC ($\bar{b}_t=0,a=0$) with  $N=10^4$.
 }
\label{polyLV}
\end{figure}

In Fig.\ref{polyLV}, we depicted the LV ratio $b$ as a function of $M_{r}$
for $n \leq 5$.  The data plots of numerical situation
are located almost everywhere
 on the constant line of $n=5$  Plummer's model  except for the outer
region. As the value of $n$ decreases, the slope
of function  $b(M_r)$ becomes steeper
 and deviates more from the LV relation (\ref{localvir2}). 

\begin{figure} 
\begin{center}
\includegraphics[width=8cm]{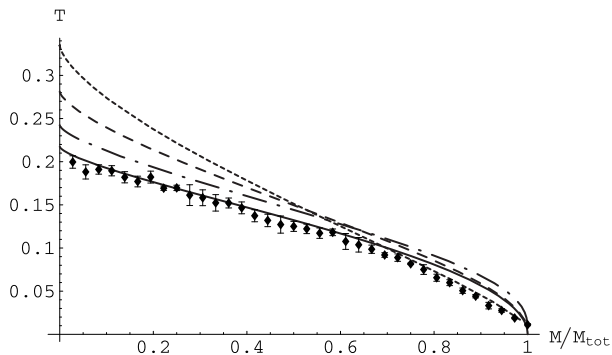}
\end{center}
\caption{ Temperature versus cumulative Mass ($M_r$) for polytropes  with the unit of $G=r_h=M_{tot}=1$. The
solid curve refers to the local temperature $k_B T/m$ for Plummer's model
(\ref{plummerTMrel}). The other curves refer to $k_B T/m$ for polytropes
with $n=3.0$ (dot-dashed), $n=1.5$ (dashed), $n=0.51$ (dotted).  The plots represent the numerical data of SC ($\bar{b}_t=0,a=0$) with  $N=10^4$.}
\label{polyTM}
\end{figure}

\begin{figure} 
\begin{center}
\includegraphics[width=8cm]{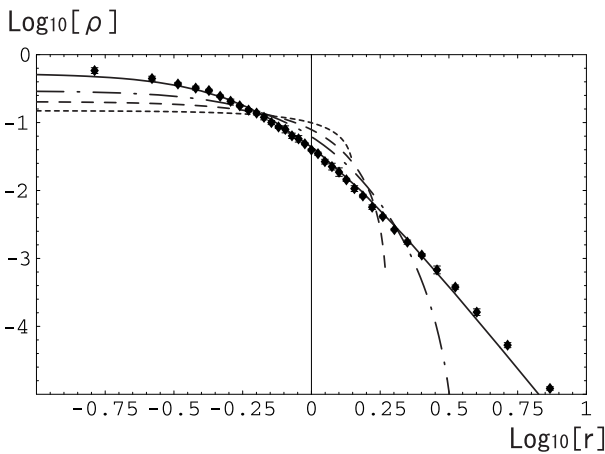}
\end{center}
\caption{ The density  for polytropes   (\ref{rho_plimmer})  with the unit of $G=r_h=M_{tot}=1$. The
solid curve refers to the density  for Plummer's model. The other curves refer to the density  for polytropes
with $n=3.0$ (dot-dashed), $n=1.5$ (dashed), $n=0.51$ (dotted). 
 The plots represent the numerical data of SC ($\bar{b}_t=0,a=0$) with  $N=10^4$.}
\label{polyrho}
\end{figure}

TM relation of Plummer's model is obtained as (\ref{plummerTMrel}) and
the slope $dT/dM_{r}$ diverges as $M_{r}\rightarrow 0$. This discrepancy is
plausible, since (\ref{sotakai}) is unphysical in the inner region. However
the deviation from the linear TM relation is conspicuous only in the central 
part of the bound state,i.e., $M_r/M_{tot}<0.05$. This is consistent
with the result that the unphysical region of (\ref{sotakai}) is at most
inner five percent of the total mass. The
slope of Plummer's model becomes almost constant and the data plots are
located almost everywhere 
in the middle region from $%
0.2M_{tot}<M_{r}<0.8M_{tot}$. In fact, the coincidence
of the  Plummer's model to 
to the numerical data is remarkable,
when  we normalize both of the physical variables with the
unit of $G=r_h=M_{tot}$, where $r_h$ is a half-mass radius of the bound system.
(Fig.\ref{polyTM} and Fig.\ref{polyrho}). 

However, the polytropes with $n \sim 5$ fail in keeping the TM relation in
the outer region since $T(M_{r})$ falls off steeply there (Fig.\ref{polyTM}%
). 
 Recalling that the constant slope of TM
relation up to the outer region is a key gradient for the DT distribution %
\cite{Osamu04}, the polytropes with $n\sim 5$ fail in reproducing the
velocity distribution characteristic after a cold collapse. From (\ref
{simple_phi}) and (\ref{phi_plimmer}), the mass density of Plummer's
solution behaves as $\propto r^{-5}$ in the outer region. This is also
inconsistent with the numerical results and with the behavior of (\ref
{sotakai}).  

\begin{figure} 
\begin{center}
\includegraphics[width=8cm]{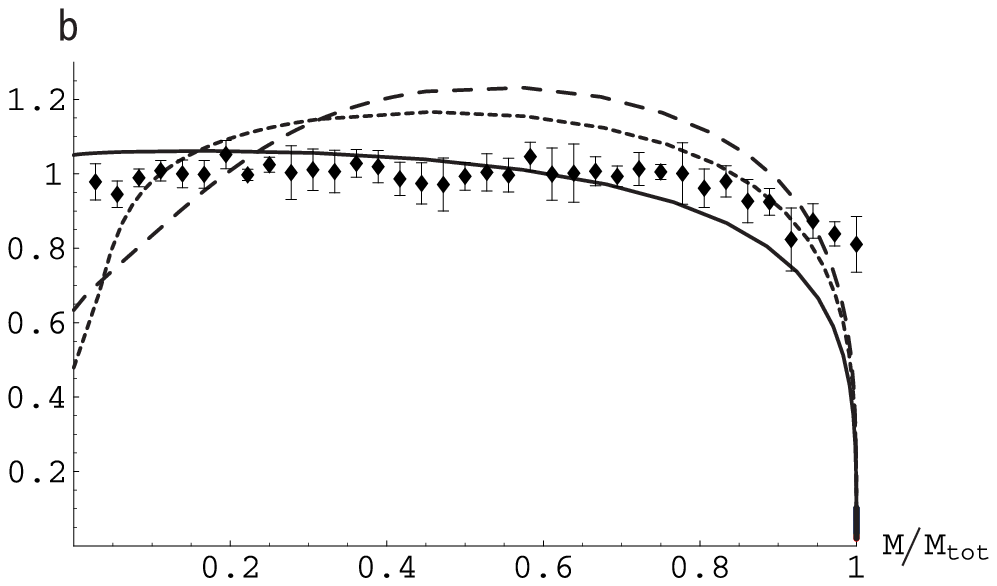}
\end{center}
\caption{ The same as Fig.\ref{polyLV} but for King models  with the unit of $G=r_h=M_{tot}=1$. The solid curve
refers to $b(M_r/M_{tot})$ for King model with $c=0.67$. The other curves refer to $%
b(M_r/M_{tot})$ for King models with $c=2.12$ (dashed), $%
c=2.72$ (dotted).  The plots represent the numerical data of SC ($\bar{b}_t=0,a=0$) with  $N=10^4$.}
\label{kingLV}
\end{figure}

\begin{figure} 
\begin{center}
\includegraphics[width=8cm]{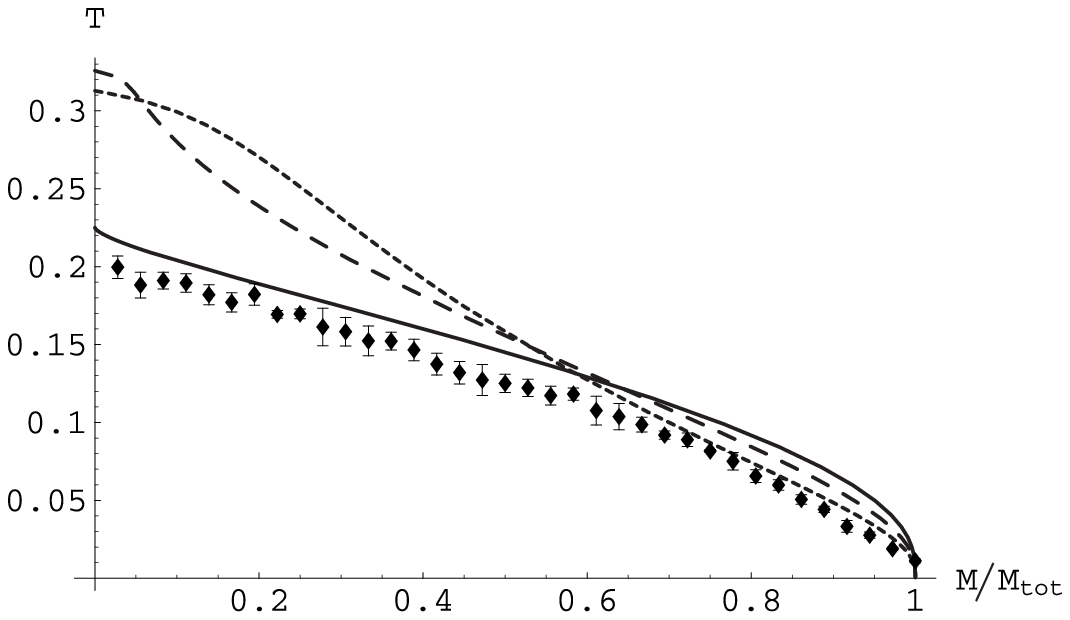}
\end{center}
\caption{ The same as Fig.\ref{polyTM} but for King models with the unit of $G=r_h=M_{tot}=1$ . The solid curve
refers to $k_B T/m$ for King model with $c=0.67$. The other curves refer to $%
k_B T/m$ for King models with $c=2.12$ (dashed), $%
c=2.72$ (dotted).   The plots represent the numerical data of SC ($\bar{b}_t=0,a=0$) with  $N=10^4$.}
\label{kingTM}
\end{figure}

Secondly, we examined King models. King models have often been applied to
fit the photometries of elliptical galaxies (e.g. %
\cite{Kormendy77,Bertin88}). They are conventionally analyzed with the
concentration parameter $c := \log_{10}(r_t/r_c)$, where $r_t$ and $r_c$ are
the tidal radius and the core radius, respectively \cite{Binney87}.  It is
shown that the King model with $c \sim 2.25$ fits the  observational $R^{1/4}
$ law quite well \cite{Kormendy77}. Here we examined the LV relation and
the TM relation for King models in the range $0.5 < c < 3.5$. For the
family of such a parameter range, $b(M_r)$ becomes smaller in the central
part for larger value of $c$, which means that the potential energy is
locally more dominated than the kinetic energy in the central part. By
contrast,  for $c \sim 0.5$, the virial ratio becomes flat close to the
center and takes the similar form to those of polytropes with $n \sim 5$(Fig.%
\ref{kingLV}). As for the TM relation for King models, $T(M_r)$ of the
numerical result fits
well to the case with $c\sim 0.5$ in the inner part $M<0.6M_{tot}$ 
but shifts to 
the case with  $c\sim2.7$ in the outer part (Fig.\ref{kingTM}).
The density profiles of King models with $c > 2.0$
 also fit  well to the numerical data in the outer part, where
it behaves as  $\rho
\sim r^{-3} - r^{-4}$.
This is consistent with the fact
that  the case with $c \sim 2.25$ fits to the  observational $R^{1/4}
$ law quite well. In contrast, the density profile for the case with  $c\sim 0.5$ 
deviates from the numerical results remarkably.

\begin{figure} 
\begin{center}
\includegraphics[width=8cm]{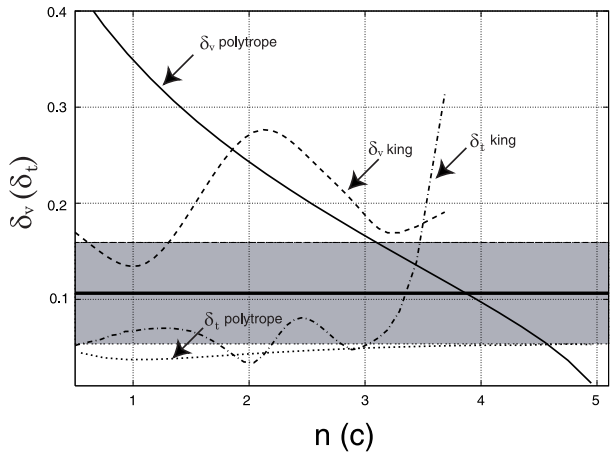}
\end{center}
\caption{Deviation $\protect\delta_v$ and $\protect\delta_t$ versus model
parameters for polytropes and King models. Residuals with respect to
polytrope index $n$ for polytropes and concentration parameter $c$ for King
models are plotted. Each curve refers to $\protect\delta_v$ for polytropes
(solid), $\protect\delta_v$ for King (dashed), $\protect\delta_t$ for
polytropes (dotted), $\protect\delta_t$ for King (dot-dashed). The gray zone
refers to the range of $\protect\delta_v=0.106\pm 0.053$ derived from the
numerical simulations of cold collapse in Fig.\ref{TM} and Fig.\ref{LV}.
Thick horizontal line denotes $\overline{\protect\delta_v}=0.106$. }
\label{TMLVfluc}
\end{figure}

Finally, we estimate the deviations from LV relation and linear TM relation
for polytropes and King models more quantitatively. Here we define the
deviations from LV relation as $\delta _{v}:=\sqrt{\left\langle \langle
(b-1)^{2}\right\rangle \rangle }$ where $\left\langle \langle \ast
\right\rangle \rangle $ denote the average of the functions of $M_{r}$
defined as 
\begin{equation}
\left\langle \langle \ast \right\rangle \rangle :=\frac{1}{{M_{tot}}}%
\int_{0}^{M_{tot}}{\;\ast \;dM_{r}}.  \label{Mav}
\end{equation}
For the definition of the deviation from linear TM relation, we first
linearly extrapolate the function $T_{e}(M_{r})$ as the form of $%
T_{e}(M_{r})=AM_{r}+B$. Then we define the deviation as 
\begin{equation}
\delta _{t}:={\frac{\sqrt{\left\langle \langle (T-T_{e})^{2}\right\rangle
\rangle }}{T_{1/2}}},
\end{equation}
where $T_{1/2}$ is the local temperature at the half mass radius. In the
parameter range we examined, we plotted both $\delta _{v}$ and $\delta _{t}$
for polytropes and King models (Fig.\ref{TMLVfluc}). We also calculated the 
$\delta _{v}$s derived from the numerical simulations of cold collapse. They
are obtained by averaging $(b(M_{r}/M_{tot})-1)^{2}$ at each $M_{r}$ in the
same way as the integral (\ref{Mav}). We got the result that the $\delta _{v}
$ derived from the simulations in Fig.\ref{TM} and Fig.\ref{LV} takes the
value $0.106\pm 0.053$, which is depicted as the gray zone in Fig.\ref
{TMLVfluc}. The parameter range whose $\delta _{v}$ exceeds this gray zone
is considered to be incompatible with the numerical simulations in Fig.\ref
{LV}.

It is obvious that the $\delta _{v}$ almost vanishes for the polytropes 
$n\sim 5$, since $n=5$ Plummer's model exactly satisfies the LV relation,
but for $n$ becomes small, $\delta _{v}$ becomes larger and exceeds the gray
zone for 
$n\lesssim 3$ 
in Fig.\ref{TMLVfluc}. Hence polytropes with 
$n \lesssim 3$ 
cannot
follow the LV relation as well as the results of numerical simulations. On
the other hand, the $\delta _{t}$s for polytropes are almost constant for
any parameter range $0.5<n<5$ and the value is almost $0.05$. As for King
models, the $\delta _{v}$ is larger than $0.1$ for all of the parameter
range and has a peak around $c=2$. Especially, the $\delta _{v}$ of 
$1.6\gtrsim c \gtrsim 2.9$ 
is larger than $0.2$, which obviously exceeds the gray zone in
Fig.\ref{TMLVfluc}. This is in contrast to the fact that the surface
brightness of King models fits well to $R^{1/4}$ law in this parameter
range. On the other hand, the $\delta _{t}$s of King models are less than $%
0.1$ except for the high concentration case $c \gtrsim 3$. The strong deviation for $%
c \gtrsim 3$ is obvious, since the central temperature becomes much higher than
those in the outer side for this parameter range.

In summary, 
we can see the remarkable difference between the polytrope sequence and the
King sequence as the collisionless static system.
In the case with King, the higher value of parameter $3.0>c>2.0$ is favorable
to the density profile or linear TM relation but is not for the LV
relation. The lower parameter range $0.5<c<1.0$ is favorable both to
the linear TM relation and to the LV relation but is not for the density
profile. Hence there is no parameter range in King
model which is favorable to  all of the characteristics 
of the bound state after a cold collapse.
On  the other hand, as for  the polytrope,
the parameter value  $n\sim5$ is favorable  to all of those
characteristics in the bound state after a collapse. 
Note that we can see the difference when we pay attention not only to 
the density profile but also to the velocity information  such as
the LV relation or linear TM relation.

Concerning the detail of the behaviour of physical quantities
for $n\sim5$ polytropes with the centrally flat boundary condition
such as Plummer's model, they fit quite well to the numerical results
in the inner region, $M_r < 0.6 M_{tot}$. In the outer region, however, both 
the density profile and the TM relation deviate from the numerical data.
We will comment on these failures of the Plummer's model or $n\sim5$ polytropes   in the 
outer part from a viewpoint of  the anisotropy of a velocity dispersion 
in the concluding remarks.



\begin{figure} 
\begin{center}
\includegraphics[width=8cm]{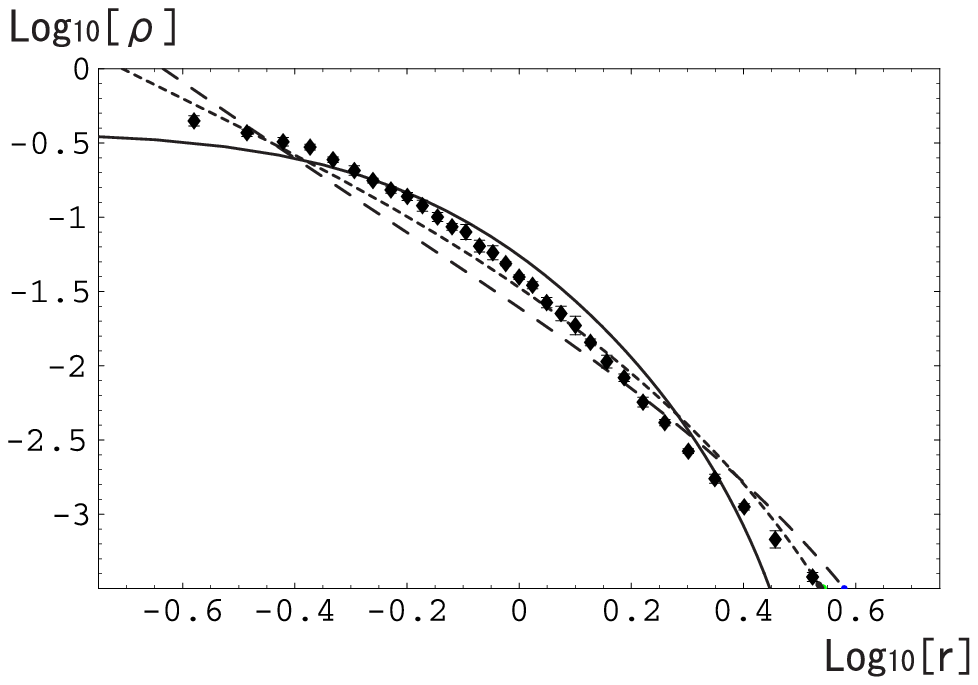}
\end{center}
\caption{ The same as Fig.\ref{polyrho} but for King models  with the unit of $G=r_h=M_{tot}=1$. The solid curve
refers to the density for King model with $c=0.67$. The other curves refer to 
the density  for King models with $c=2.12$ (dashed), $%
c=2.72$ (dotted).  The plots represent the numerical data of SC ($\bar{b}_t=0,a=0$) with  $N=10^4$. }
\label{kingrho}
\end{figure}

\section{Concluding remarks}

\label{sec:conclusions} In this paper, we have proposed (a) the linear TM
relation and (b) the LV relation as the hypotheses of self-gravitating
structures after a cold collapse. Based on these two hypothetical proposals,
we have uniquely determined the mass density profile (\ref{sotakai}) in the
quasi-equilibrium state.

Our density profile behaves as $\rho \sim r^{-4}$ in the outer region
and fits well both to the data of cold collapse simulations.
Although this density profile is unphysical in the inner region, 
where it decreases with the decreasing radius,
we showed that such an unphysical region is just a few percent 
around the central region of the bound state. Hence,
we can safely neglect this region at least for the resolution
of our numerical results. 

In the central region, we also have to notice the
relaxation process in such a central region.
In fact, we examined how two-body relaxation is developed 
in a  bound state after a cold collapse. 
We estimated the time scale of the  two-body
relaxation with $t^{\ast}_{rel}$, which is defined as  
the time when the region within a half-mass radius
achieves a two-body
relaxation on average.
Both the linear TM relation and the LV relation appear 
even when the two-body relaxation is not attained.
Hence, we can say that our two hypotheses are not affected through
the two-body relaxation but are characteristic even in collisionless system
in our estimation with $t^{\ast}_{rel}$. 

From a viewpoint of these two hypotheses, we compared our results with two
isotropic static models,i.e., polytropes and King models. The polytrope with $n=5$ is
special among those models, since it exactly satisfies the LV relation.
Plummer's model, the polytrope with $n = 5$ with a flat boundary 
condition at the center 
 also satisfies the constant TM
relation quite well in the intermediate region $0.2M_{tot}<M_{r}<0.8M_{tot}$. 
Moreover, its density fits well
to the numerical data quite well in this intermediate region. On the other hand, both
its TM relation and density profile in the outer region are slightly shifted  from those
 in cold collapse simulations: the temperature falls off more
steeply and the density behaves as $r^{-5}$ in the outer region. 
As for King models, in contrast, there is no parameter range in which both the TM relation
and density profile fit well to the numerical data even in the intermediate
region. 

We also examined the deviations from the two hypotheses quantitatively for
these two models. Only the polytropes with $n \sim 5$ are acceptable in that their
deviations from both of the hypotheses are compatible with the numerical
results of cold collapse. The LV relation imposes a strong constraint for
these models. Especially for King models, the deviation from LV relation
takes the maximum value around the parameter $c\sim 2$, in which the surface
brightness fits well to $R^{1/4}$ law. This suggests that considering only
the asymptotic behavior of the bound state is not enough to describe the
collisionless static models to express observational or numerical results.

Combining the results for the above two models, we can speculate
that Plummer's model follows both of the hypotheses  we proposed (a) LV
relation and (b) linear TM relation. However,
this model is  lack of the
following properties of the bound states in the actual simulations of cold
collapse: (i) deviation from a hydrostatic equilibrium state and (ii)
anisotropy of velocity dispersion in the outer region.

As for (i), the hydrostatic equilibrium condition is not attained especially
when the violent degree of the initial collapse is strong, since the bound
state slowly expands outward and is not balanced with a gravitational force
after a cold collapse. Hence, to express these conditions we may have to
consider not the static solution but the stationary solution admitting the
coherent motion of matters.

As for (ii), we derived that for 
a mild collapsing case, the LV relation fluctuates more largely
in the outer region and the slope of the TM relation
 at this region becomes a little
steeper at least in the collisionless stage.
We suspect that these results are related to the
anisotropy of this system.  Hence for more precise description
of the system possessing two hypotheses, it seems necessary
to treat the model with the anisotropy of the velocity dispersion.
So far  several collisionless static models with the anisotropic
velocity dispersions or with the non-spherical symmetry have been proposed
 \cite{Henon73,Evans05,Bertin84,Stiavelli85,Dejonghe87,Gerhard91,Bertin03}.
Incompleteness of the violent relaxation or the existence of conservative
quantities during the phase mixing seem to be a candidate as the model to 
induce such anisotropy \cite{Lynden67,Stiavelli85}. 
The hypotheses we proposed are expected to be used as a guideline to
build up the models following the characters of the bound states after a
cold collapse by combining with these anisotropic models.

It also seems important how widely our two hypotheses are applicable in more
general mixing processes including a merging process which yields a cuspy
density profile in the central region \cite{Navarro96}. If we want to know the detail of 
such a tiny central region, such as the existence of 
the cuspy density profile, we have to treat  such a
central region more carefully. We speculate that in such a region, at least
one of the two hypotheses we proposed will be broken. 
These points are
now under investigation.

\section*{Acknowledgements}

We would like to thank Prof. Kei-ichi Maeda and Prof. Stefano Ruffo for the
extensive discussions. All numerical simulations were carried out on GRAPE
system at ADAC (the Astronomical Data Analysis Center) of the National
Astronomical Observatory, Japan.

\appendix

\section{Mass fraction of the centrally flat region against the outer analytical solution (\ref{sotakai}) }

\label{appjoin}

The analytical solution (\ref{sotakai}) has an inner unphysical region at $r <r_{\ast}$. Since
the density goes to zero toward the center in this region, the cumulative mass in this region
becomes very small. Here we will check how much the value is modified when we put the constant
mass density in this region and connect it to the analytical solution at $r =r_{\ast}$.
From the continuity at  $r =r_{\ast}$, the density becomes 
\be
\rho  = \left\{ {\begin{array}{*{20}c}
   {\rho _0 \left( {{4 \mathord{\left/
 {\vphantom {4 e}} \right.
 \kern-\nulldelimiterspace} e}} \right)^4 } \hfill & {\left( {r < r_ *  } \right)} \hfill  \\
   {\rho _0 \left( {{{r_0 } \mathord{\left/
 {\vphantom {{r_0 } r}} \right.
 \kern-\nulldelimiterspace} r}} \right)^4 e^{ - {{r_0 } \mathord{\left/
 {\vphantom {{r_0 } r}} \right.
 \kern-\nulldelimiterspace} r}} } \hfill & {\left( {r \ge r_ *  } \right)} \hfill  \\
\end{array}} \right.
.
\ee

The cumulative mass at $r =r_{\ast}$ is
\be
M_{in}  = 4\pi \rho _0 r_0^3 \left( {\frac{4}{{3e^2 }}} \right).
\ee
The cumulative mass outside  $r_{\ast}$ is
\be
M_{out}  = 4\pi \int_{{{r_0 } \mathord{\left/
 {\vphantom {{r_0 } 4}} \right.
 \kern-\nulldelimiterspace} 4}}^\infty  {\rho _0 \left( {\frac{{r_{_0 } }}{r}} \right)^4 e^{ - {{r_0 } \mathord{\left/
 {\vphantom {{r_0 } r}} \right.
 \kern-\nulldelimiterspace} r}} } r^2 dr = 4\pi \rho _0 r_0^3 \left( {1 - e^{ - 4} } \right).
\ee
Hence the mass fraction of the inner region becomes
\be
\frac{{M_{in} }}{{M_{tot} }} = \frac{{M_{in} }}{{M_{in}  + M_{out} }} = \frac{4}{{3e^4  + 1}} \approx 0.0243.
\ee
\section{Derivation of the  LV relation for polytropes}

\label{ap-LV} \vspace{1cm} Following \cite{Binney87}, we use the notation
using the relative energy $\epsilon$ and relative potential $\phi$ defined
as 
\begin{eqnarray}
\tilde{E} &:=& -E+\Phi_{\ast},  \nonumber \\
\phi &:=& -\Phi+\Phi_{\ast},  \label{relativedef}
\end{eqnarray}
respectively. For polytropes, using these notations, the distribution
function is described as 
\begin{equation}
f(\tilde{E}) = \left\{ {\begin{array}{*{20}l} \kappa_1 \tilde{E}^{n-3/2},
\;\;\;&\;(\tilde{E} \geq 0); \\ 0, \;\;\;&\;(\tilde{E} < 0), \\ \end{array}}
\right.  \label{deff}
\end{equation}
For (\ref{deff}), density $\rho$ and temperature $T$ are simply expressed as 
\begin{eqnarray}
& &\rho= c_n \phi^{n},  \label{simple_phi} \\
& &<v^2>={\frac{3k_BT }{m}}={\frac{3 }{n+1}} \phi,  \label{simple_T}
\end{eqnarray}
where $c_n := (2\pi)^{3/2}{\frac{(n-3/2)! }{n!}}%
\kappa_1$.

From (\ref{poteall}) and (\ref{relativedef}), we get
\begin{equation}
\overline{W}_r={\frac{M_r \Phi_{\ast} }{2}}-2\pi \int_0^{R} \phi \rho(r) r^2
dr.  \label{poteall2}
\end{equation}
From (\ref{kineall}),(\ref{poteall2}), and (\ref{simple_T}), we get
\begin{equation}
\overline{K}_r + {\frac{3 }{n+1}}\overline{W}_r={\frac{3 }{2(n+1)}}M_r
\Phi_{\ast}.
\end{equation}
Especially for $n=5$, from the condition $\Phi_{\ast}=0$, the LV relation $2%
\overline{K}_r/|\overline{W}_r|=1$ is obtained.

\section{Derivation of TM relation for Plummer's model}

Following \cite{Binney87}, we define the dimensionless variables

\begin{equation}
\psi :=\phi /\phi _{0}\;\;\;\;s:=r/b,  \label{deffpoly}
\end{equation}
where $\phi _{0}$ is the value of the relative potential at the center and $%
b:=\{4\pi Gc_{n}\phi _{0}^{n-1}\}^{-1/2}$. Then, $\psi $ is followed by
Lane-Emden equation, 
\begin{equation}
{\frac{1}{s^{2}}}{\frac{d}{ds}}\left(s^{2}{\frac{d\psi }{ds}}\right)=\left\{ {%
\begin{array}{*{20}l} -\psi^n, \;\;\;&\;(\psi \geq 0); \\ 0, \;\;\;&\;(\psi
\leq 0), \\ \end{array}}\right.   \label{polyeqk}
\end{equation}
with the boundary condition 
\begin{equation}
\lim_{s\rightarrow \infty }\psi (s)=\psi _{\ast },  \label{boundaryp3}
\end{equation}
where $\psi _{\ast }:=\Phi _{\ast }/\phi _{0}$. Especially for $n=5$, the
solution of (\ref{polyeqk}) with the flat boundary condition at the center
becomes 
\begin{equation}
\psi ={\frac{1}{\sqrt{1+s^{2}/3}}},  \label{phi_plimmer}
\end{equation}
and meets the boundary condition 
\begin{equation}
\lim_{s\rightarrow \infty }\psi (s)=\psi _{\ast }=0.  \label{boundaryplum}
\end{equation}
The density is described as
\bea
\rho=\psi^n=\frac{1}{(1+s^{2}/3)^{n/2}}. \label{rho_plimmer}
\eea

In spherically symmetric case, the cumulative mass $M_{r}$ is expressed as $%
M_{r}={\frac{r^{2}}{G}}{\frac{d\Phi }{dr}}$, whose dimensionless form
becomes 
\begin{equation}
M_{r}=-s^{2}{\frac{d\psi }{ds}},  \label{mphi}
\end{equation}
and the total mass $M_{tot}$ is equal to $\sqrt{3}$. Substituting (\ref
{phi_plimmer}) into dimensionless form of (\ref{simple_T}) and (\ref{mphi})
and eliminating $s$, we can easily derive the TM relation of Plummer's
model as 
\begin{equation}
{\frac{k_{B}T}{m}}={\frac{1}{6}}\sqrt{1-\left( {\frac{M_{r}}{M_{tot}}}%
\right) ^{2/3}}.  \label{plummerTMrel}
\end{equation}




\end{document}